\documentclass[a4paper,11pt]{article}
\usepackage{jinstpub} % for details on the use of the package, please see the JINST-author-manual
\usepackage{lineno}
\usepackage{color}
\usepackage{siunitx}
\usepackage{graphicx}
\usepackage{float}
\usepackage{pdfpages}
\usepackage{appendix}
\usepackage{anyfontsize}
\usepackage{amsmath}
\usepackage{hyperref}

\DeclareSIUnit\bar{bar}
\DeclareSIUnit\atomicmassunit{u}

\title{Development and commissioning of ion-optical elements for ion and antiproton beams with energies up to 5\,keV}

\author[a,b]{C. Klink}
\author[a]{M. Schlaich}
\author[a]{J. Fischer}
\author[a]{A. Obertelli}
\author[a]{A. Schmidt}
\author[a]{F. Wienholtz}

\affiliation[a]{TU Darmstadt, Schlossgartenstraße 9, 64289 Darmstadt, Germany}
\affiliation[b]{CERN, Genève, Switzerland}

\emailAdd{clara.klink@cern.ch}

\abstract{In nuclear and atomic physics experiments, charged ion beams often need to be guided from the ion production to the experimental site. In the PUMA experiment, an ion source beamline was developed, which can be operated with up to \SI{5}{\kilo\electronvolt} beam energy at a base pressure of $10^{-9}$\,mbar or better. In this paper, a low-energy pulsed drift tube for beam energy modification, a hybrid einzel lens assembly for beam focusing and steering and an iris shutter assembly for separating beamline sections with different vacuum requirements are described with their design principles and performances.}

\notoc
\begin{document}
\maketitle
\flushbottom
\section{Introduction}
The PUMA (antiProton Unstable Matter Annihilation) experiment aims at studying the neutron-to-proton ratio in the density tail of stable and exotic nuclei, indicative of nuclear phenomena like halo nuclei and neutron skins by use of low-energy antiprotons \cite{Aumann2022, Oliver:2847847}. The first goal of the experiment is to characterize the neutron-to-proton ratio of stable isotopes at the Antiproton Decelerator (AD) at CERN in a Penning-Malmberg trap by mixing them with antiprotons. The ions of interest have to obey strict requirements with respect to the beam emittance to allow for an efficient injection into the PUMA Penning trap, and to the isotopic purity for the subsequent annihilations with antiprotons. Thus, a dedicated, versatile ion source beamline was designed to create, transport and shape the beam from the ion source to the PUMA Penning trap. \\
\section{Experimental setup}
\begin{figure}
    \centering
   \includegraphics[width=1\textwidth]{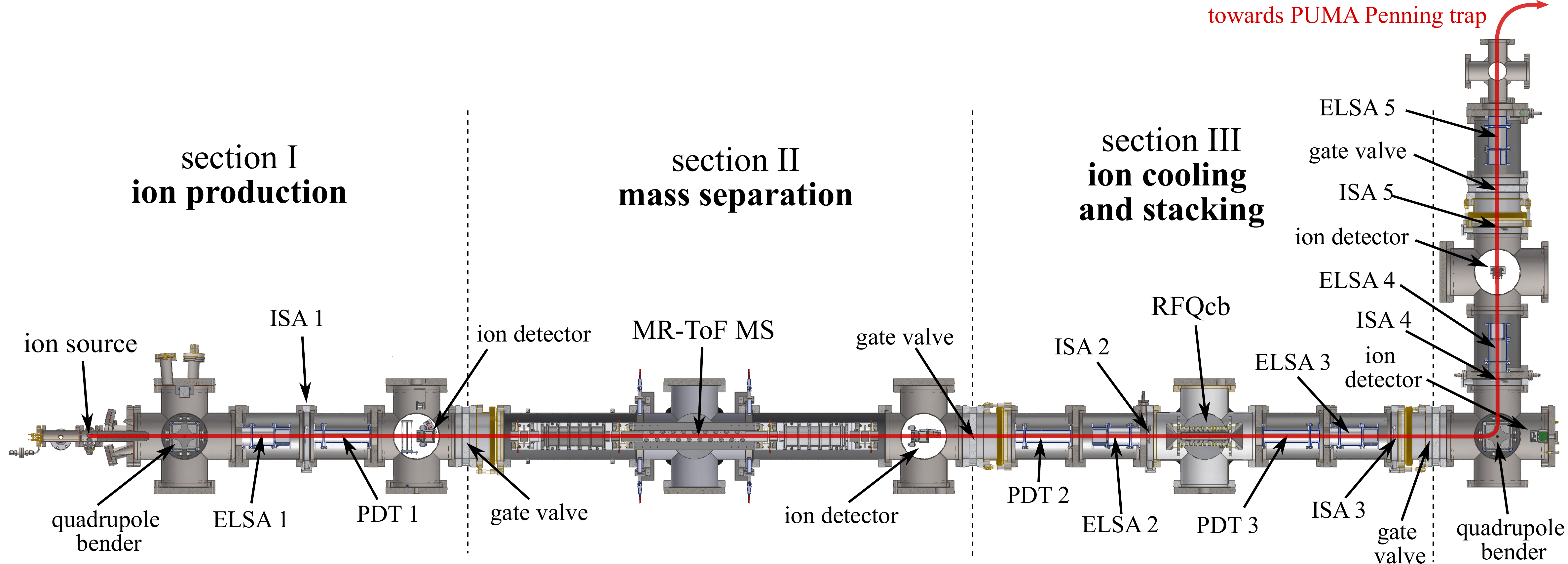}
    \caption{The ion source beamline of the PUMA experiment consists of four sections: ion production, mass separation, ion cooling and stacking and transfer to the antiproton beamline. The beamline includes einzel lens and steerer assemblies (ELSA), pulsed drift tubes (PDT) and iris shutter apertures (IS). After preparation, the beam is guided towards the PUMA Penning trap.}
    \label{fig:beamline_ion_source}
\end{figure}

In figure \ref{fig:beamline_ion_source}, a schematic view of the beamline is shown. The ions are produced in bunches with energies of up to \SI{5}{\kilo\electronvolt} in an electron impact plasma ion source (IQE 12/38, SPECS Surface Nano Analysis GmbH). The beam is then purified in a multi-reflection time-of-flight mass separator (MR-ToF MS), capable of achieving a mass resolving power $m$/$\Delta m$ of approximately \,10$^5$ \cite{SCHLAICH2024117166}. Further downstream, the bunches are cooled and accumulated in a quadrupole radio-frequency cooler-buncher (RFQcb) \cite{LECHNER2024169471} which uses the buffer-gas cooling technique \cite{KELLERBAUER2001276}. \\
During normal operation, the desired ion species is produced with a kinetic energy of 3 keV. The most critical part in terms of beam transportation is the injection and ejection of the beam into and out of the RFQcb, whose injection cone has an inner diameter of \SI{2}{\milli\meter}. Furthermore, for a successful trapping in the biased RFQcb, the ions have to be decelerated from an energy of about \SI{3}{\kilo\electronvolt} to \SI{0.3}{\kilo\electronvolt} while keeping the beam spot below \SI{2}{\milli\meter}. \\ %The components have to be designed considering the longitudinal and transversal emittance of the beam, which is estimated to be \textcolor{red}{emittance} determined with \textcolor{red}{name + citation of pepper pot}. \\
In order to meet these requirements, dedicated ion-optical elements were designed to be installed in between the sections of the beamline: an electrostatic einzel lens in combination with a segmented steerer for steering and focusing, as well as a pulsed drift tube (PDT) for modifying the kinetic energy of the ion beam. Additionally, the strict vacuum requirements (UHV,  < $10^{-9}$\SI{}{\milli\bar}) at the end of the ion transfer section have to be respected. Thus, a separation of the vacuum sections with adjustable apertures based on iris shutters was implemented. In this work, the three beamline elements are presented in detail.

\section{Simulation method}
The design of the ion-optical elements was determined based on simulations with SIMION\textsuperscript{\textregistered}2020, a software package used to calculate electric and magnetic fields based on a given configuration of electrodes and magnets by solving the Laplace equation \cite{Simion:manual}. Extensive descriptions on the simulation methods used in SIMION\textsuperscript{\textregistered}  can be found, e.g., in reference \cite{SIMION_Soliman}. All electrodes were defined in a \SI{1}{\milli\meter} grid and refined with a convergence objective of $10^{-7}$ \SI{}{\volt}.\\
In the simulation, $^{40}$Ar$^+$ ions were considered following a Gaussian angular distribution with a FWHM of \SI{1.5}{\degree} and a Gaussian spatial distribution of \SI{1}{\milli\meter}, corresponding to a transverse FWHM emittance of 8.3\,$\pi$\,mm\,mrad, an initial kinetic energy of \SI{3000}{}\,$\pm$\SI{10}{\electronvolt} FWHM and a uniformally distributed bunch length of \SI{300}{\nano\second}, which corresponds to a length of \SI{3.61}{\centi\meter} for the initial kinetic energy. The ions were created directly after the ion source.
\section{Einzel lens and steerer assembly}
\subsection{Description}
\begin{figure}
    \centering
    \includegraphics[width=1\textwidth]{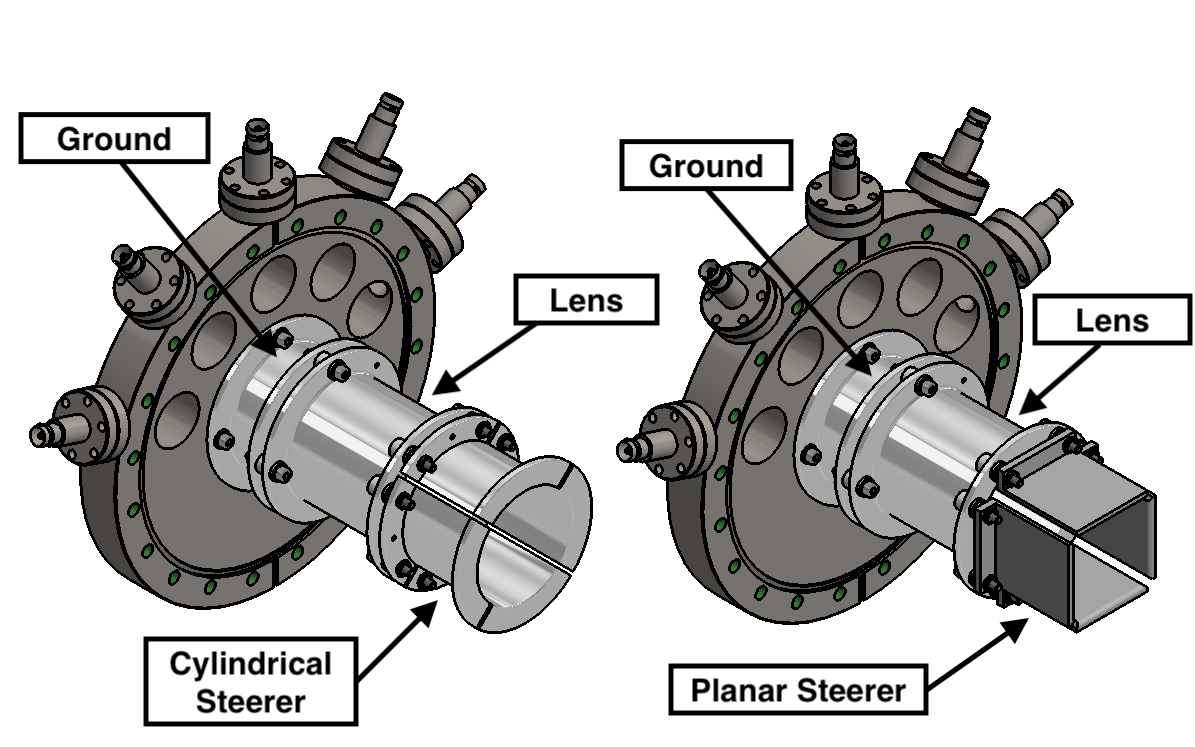}
    \caption{The einzel lens is mounted on a customized double sided CF-160 flange and consists of 6 electrodes: a grounding element, a focus element and a four-fold segmented electrode for x-y steering of the beam. Two configurations of steerer plates were tested: cylindrical (left) and planar (right).}
    \label{fig:steeres}
\end{figure}

The general design strategy for einzel lenses is, e.g., described in reference \cite{SISE2005114}. The einzel lens and steerer assembly (ELSA) is shown in Figure \ref{fig:steeres}. It is contained in a CF160 tube of length \SI{20.32}{\centi\meter}. Drawings and dimensions are provided in appendix \ref{drw:lens}. The assembly consists of three highly polished aluminum cylinders. Aluminum was chosen as material due to its low weight, cost and low outgassing rates. The lens electrode is sandwiched between a four-fold segmented steerer electrode and a grounded electrode. The grounded electrode is connected directly to the flange. The steerer electrodes are separated from the focus electrode by alumina (Al$_2$O$_3$) split bushes (tectra, BGB-M4) to be able to supply high voltage to all electrodes. Their assembly is also shown in appendix \ref{drw:lens}. The maximum applied voltage to the electrodes are $\pm$ \SI{5}{\kilo\volt} DC, limited by the electrical feedthroughs. The ELSA is mounted on a customised double-sided CF160 flange, with five radially positioned CF-16 flanges, that are equipped with single-pin SHV-5 feedthroughs each (see appendix \ref{drw:flange}). Two geometries are considered for the assembly, which are shown in figure \ref{fig:steeres}, one with cylindrical and one with planar steerer electrodes. The ELSA is used successfully as a standard component in the PUMA ion source and antiproton beamline.
%The reference electrode is grounded and defines the field lines to the focus element. Our design includes an additional four-fold segmented electrode to allow for x-y-steering, which is mounted on the focus electrode. 
\subsection{Length of focus electrode and radius of einzel lens assembly}
\begin{figure}
    \centering
    \includegraphics[width=1\textwidth]{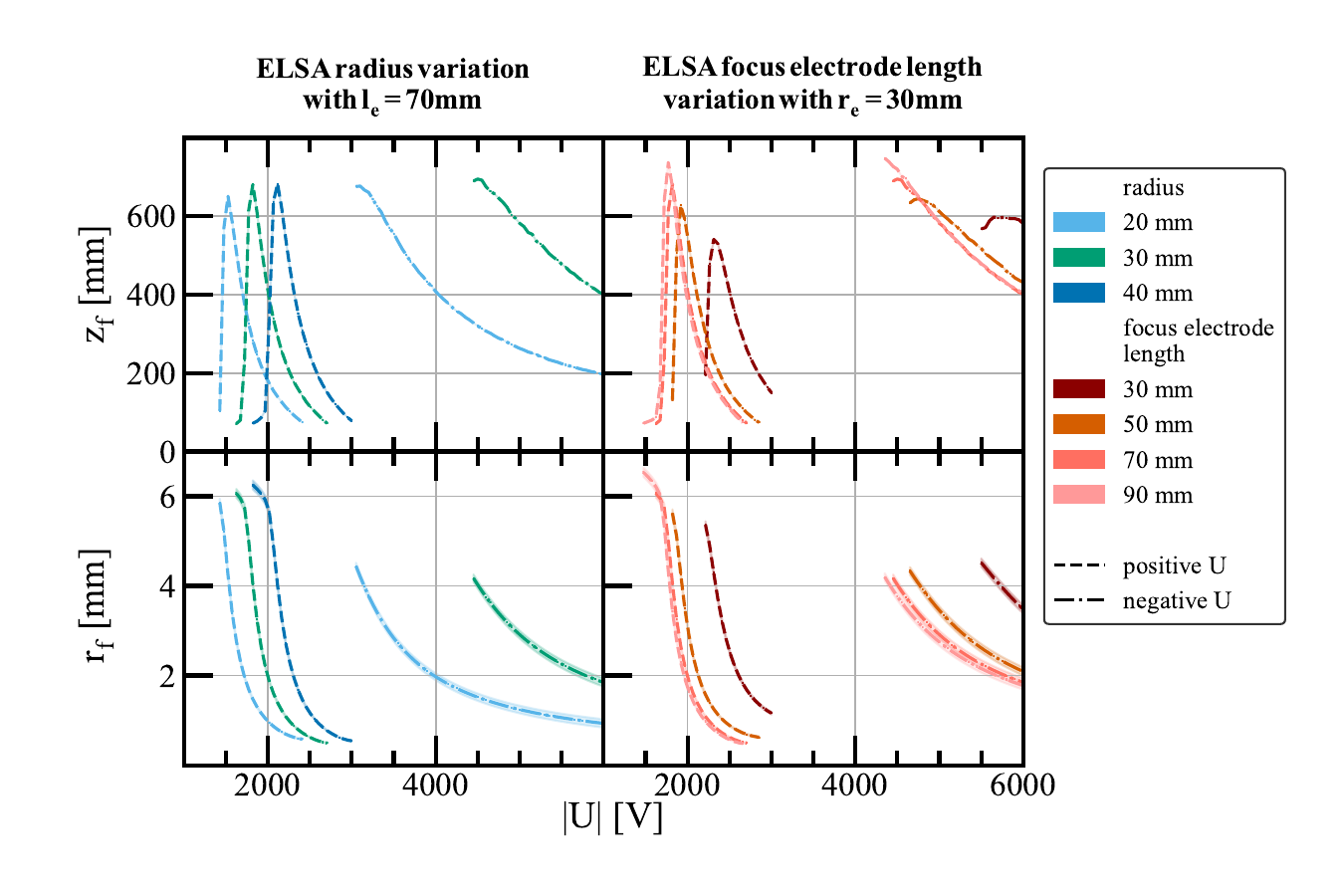}
    \caption{The length of the focus electrode $l_e$ and the radius of the lens assembly $r_e$ were varied in dependence of the applied focus voltage $U$ (abscissa) in simulations to determine the impact on the focus point properties $z_f$ and $r_f$. Positive and negative applied voltages are displayed on the same positive axis, indicated by different line styles. The dashed lines indicate a positive applied voltage, the dashed-dot lines a negative applied voltage.}
    \label{fig:geometry}
\end{figure}

The focal point position $z_f$ and beam spot size $r_f$ were determined in dependence of the applied voltage and the design parameters of the einzel lens assembly. $z_f$ is measured from the center of the einzel lens segment. The beam spot $r_f$ is the mean radius of the ion beam at $z_f$. Using only DC fields, the results for $z_f$ and $r_f$ can be scaled for different initial beam energies, thus simulations were only performed for one initial beam energy. \\
Figure \ref{fig:geometry} shows $z_f$ and $r_f$ in dependence of the applied voltage $U$ for various radii $r_e$ of the ELSA and lengths of the lens electrode $l_e$. A smaller $r_e$ allows to reach smaller $r_f$ for lower applied voltages. $z_f$ and $r_f$ saturate for $l_e$ > \SI{70}{\milli\meter}, which is short enough to contain the ELSA in the CF160 tube. The larger $r_e$, the higher is the required potential to reach small $r_f$. For ELSAs used in the case of the PUMA ion source beam line, a radius of \SI{30}{\milli\meter} was chosen. This will allow to sufficiently focus the beam to pass orifices < \SI{2}{\milli\meter} in the beam line with a $z_f$ that corresponds to the distance between ELSA and orifices, using an applied voltage of approximately \SI{2}{\kilo\volt}.
\subsection{Steerer design}
\begin{figure}
    \centering
    \includegraphics[width=1\textwidth]{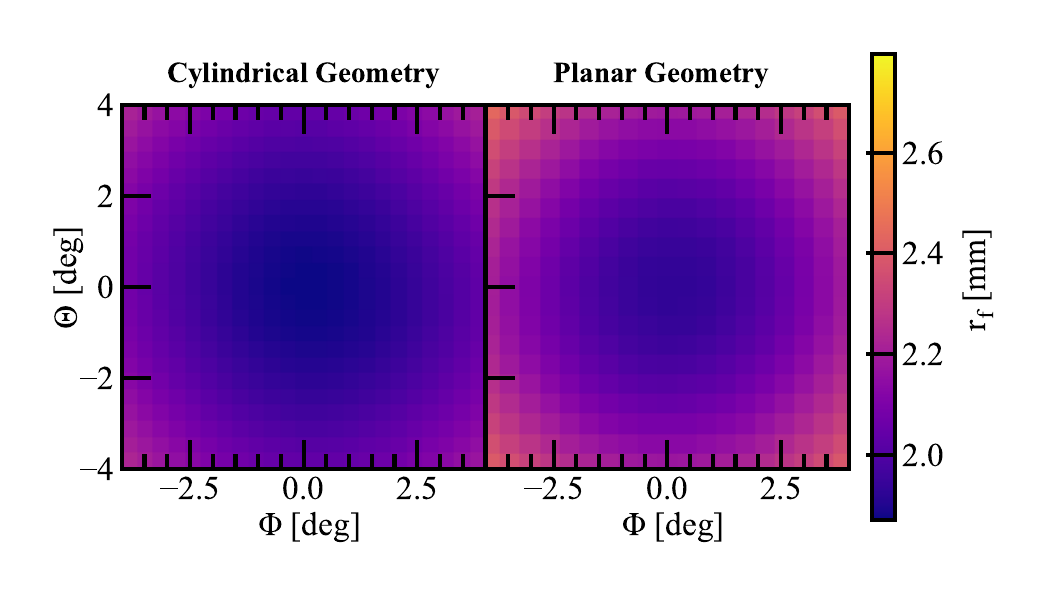}
    \caption{Simulated $r_f$ for steering angles $\Phi$, $\Theta$ for (left) the cylindrical geometry and (right) the planar geometry for an applied focus voltage $U$\,=\,\SI{2}{\kilo\volt}. $r_f$ is larger when steering the beam towards the edges of the steerers, and the effect is pronounced for the planar geometry.}
    \label{fig:steering_focus_point}
\end{figure}

The effect of the steerer shape was simulated for a cylindrical and planar geometry. A steering voltage $\pm V_s$/2 is applied to opposite steering electrodes, leading to a voltage difference $V_s$. The achieved angle in the horizontal or vertical plane was plotted for the applied voltage $V_s$. The results were fitted with a straight line:
\begin{linenomath*}
\begin{align}
    \Theta_{\text{cyl}} (V_s) = (0.0074 / \text{V} \cdot V_s - 0.0034)\,\text{°}, \\
    \Theta_{\text{pla}} (V_s) = (0.0091 / \text{V} \cdot V_s - 0.0100) \,\text{°}.
\end{align}
\end{linenomath*}
The planar geometry requires approximately 0.81 $V_s$ to reach the same steering angles as the cylindrical geometry. In the case of the PUMA ion source beamline, a maximum $V_s$ of $\pm$\SI{1}{\kilo\volt} can be applied, limited by the power supplies and cables used, leading to elevation and azimuthal angle $\Theta_{\text{cyl,max}}\approx$ \SI{7.4}{\degree} and $\Theta_{\text{pla,max}}\approx$ \SI{9.1}{\degree}. \\
The value of $r_f$ increases for a deflected beam compared to a non-deflected beam. This is furthermore influenced by the applied focus voltage $U$, e.g., when applying $U$\,=\,\SI{1.8}{\kilo\volt} in the planar steerer assembly, $r_f$ increases by 34.8\,\%, while for $U$\,=\,\SI{2.0}{\kilo\volt}, $r_f$ increases by 45.1\,\%. Further increasing $U$ to \SI{2.2}{\kilo\volt} yields an increase of 21.2\,\%. The cylindrical design has a better $r_f$ stability than the planar design, as can be seen in Fig. \ref{fig:steering_focus_point}, especially when focusing diagonally. 
\subsection{Angular deviation of the beam}
\begin{figure}
    \centering
    \includegraphics[width=1\textwidth]{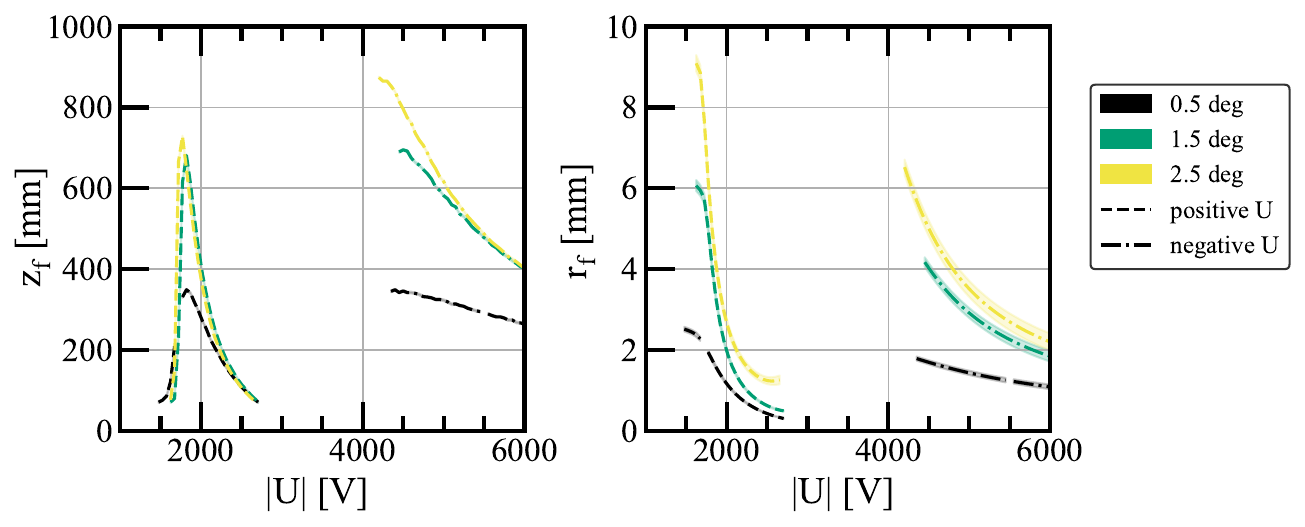}
    \caption{Simulated focal point properties for different FWHM angular deviations of the beam cone for the PUMA ELSA in the cylindrical geometry. The larger the angular deviations, the larger is the minimum $r_f$ that can be reached and the longer is $z_f$. The dashed lines indicate a positive applied voltage, the dashed-dot lines a negative applied voltage to the focus electrode.}
    \label{fig:angular}
\end{figure}

The angular deviation describes the opening half-angle of the cone in which the ions are distributed around the central trajectory. It counteracts the focusing power of the einzel lens. Thus, $z_f$ is increasing for higher angular deviations for a fixed applied potential $U$, see figure \ref{fig:angular}. Additionally, $r_f$ degrades with higher angular deviation, which will lead to beam losses in low acceptance components. In case of inefficient transmission through a beam line, additional focusing elements might be necessary to compensate for a large angular deviation. \\
The angular deviation also has an effect on the behaviour of the beam after steering. Again, the planar geometry is more affected by an increase in angular beam deviation and will lead to a bigger $r_f$ compared to the cylindrical geometry. For this reason, the cylindrical geometry was chosen for the ELSAs in the PUMA ion source beamline. 
%\subsubsection{Energy}
%In Fig. \ref{fig:lens_energy} can be seen how ions with different energies were focused with the einzel lens. For a better comparability the axis is scaled to the percentage of the ion energy. It can be concluded that there is a negligible energy-dependence of the focusing effect of the lens.
%\begin{figure}
%    \centering
%     \subfloat[Focal point position.]  {\includegraphics[width=0.49\textwidth]{Figures/Argon40_energy_zpos_long.png}\label{fig:zf_angular}}
%    \hfill
%    \subfloat[Focal point radius.]  {\includegraphics[width=0.49\textwidth]{Figures/Argon40_energy_radius.png}\label{fig:radius_angular}}
%    \caption{Focal point characteristics for beam energies. The applied voltage is shown in percentage of the beam energy.}
%    \label{fig:lens_energy}
%\end{figure}
\section{Pulsed drift tube}
\subsection{Description}
Pulsed drift tubes have been designed for a large spectrum of applications, the earliest included drift tube acceleration of heavy ions \cite{Faltens:1107958}. In more recent applications, the concept of a pulsed drift tube is used, e.g., for the capture of ions in MR-ToF devices with an in-trap lift \cite{KNAUER201746}, bunch re-acceleration after an RFQcb \cite{HERFURTH2001254} or the HV deceleration of antiprotons for the injection in particle traps \cite{FISCHER2024165318, HUSSON2021165245}. The dimensions of the low-energy PDT can be found in appendix \ref{drw:pdt}. The PDT assembly is mounted on the same electric feedthrough flange as the einzel lens and is contained in a CF160 tube. It consists of two cylindrical electrodes: a drift tube and an isolated electrode which are made of highly polished aluminum and separated by alumina split bushes. The drift tube is also separated with split bushes from the mounting flange to create a second gap between flange and drift tube. Thus, the drift segment is sandwiched between two isolated segments, which are usually grounded. An additional focusing effect can be achieved by applying high voltage to the isolated electrode.  
\subsection{Operation principle}
A pulsed drift tube modifies the kinetic energy of traversing ions. A de- or accelerating field is created in the gap between the first grounded electrode and the drift tube. After the energy modification, the ions enter an ideally field-free drift region. While the ion bunch is fully contained in the tube, the applied voltage on the tube is switched to a lower or higher value respectively. Thus, upon leaving the PDT, the ions remain in their modified energy state. In the most simple case, the drift tube is switched to or from ground respectively, but other switch patterns can be implemented. \\
Since the phase-space volume of the ion bunch is constant, decelerating the ions results in an increased spatial spread that has to be compensated by the use of ion-optical elements like an einzel lens. An acceleration will lead to a focusing effect. \\
A PDT can be characterized by relating the switch time $t_s$ to the time-of-flight (ToF). If switched when the ion bunch is outside the drift tube, the energy state of the ions will remain unchanged after traversing the PDT. If the ion bunch is only partly contained in the tube, it will also only be partly affected by the change of field leading to a high energy spread of the resulting bunch. Only if the bunch is fully contained in the tube during switching, all ions can achieve the desired energy state. The longitudinal acceptance of the drift tube has to be appropriately chosen to match the bunch length of the incoming beam. \\ 
Accordingly to the description above, also the second gap after the drift electrode can be used for energy modifications. For a larger longitudinal acceptance, the bunch should traverse the tube always in the lower-energy state. Thus, the gap before the drift tube will be used for a deceleration and the gap after the drift tube for an acceleration. Other switch patterns could be implemented, e.g., switching to higher or lower values than ground respectively, which can have an impact on the transversal and longitudinal emittance.
\subsection{Calculation of the ideal switch time}
\begin{figure}
    \centering
    \includegraphics[width=0.7\textwidth]{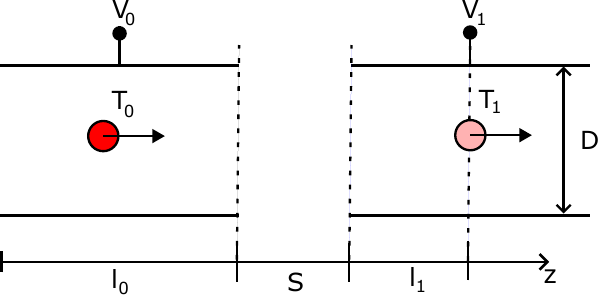}
    \caption{The ideal switch time can be calculated dependent on the dimensions and applied voltages to the PDT.}
    \label{fig:sketch}
\end{figure}

Consider a low-energy ion of mass $m$ and charge $q$ passing through a conductive concentric-to-axis cylinder with a gap of width $S$ and inner diameter $D$ with a kinetic energy $T_0$ in positive direction $z$. The first side of the cylinder has an applied voltage of $V_0$ and length $l_0$, the other side $V_1$ and 2$l_{1}$, see figure \ref{fig:sketch}. Traversing the gap will modify the particles energy to $T_1$. The ideal switch time, when the bunch is centered in the tube at $l_1$, is calculated for two different assumptions of the field in the energy modification gap.
\subsubsection{Plate capacitor}
A step-wise calculation is performed for the drift time of the ions: first, until they reach the gap, second, passing through the gap and third, from the start of the drift tube to it's center. In a simplified view, an electric field of a plate capacitor is assumed inside the gap, the region before and after the gap is field-free. Thus, ions will drift for distances $l_0$ and $l_1$ with constant velocity outside the gap, while the electric field $E$ in the gap and consequently the constant acceleration $a$ along the z-axis compute to
\begin{linenomath*}
\begin{align}
    \Vec{E}  = \frac{V_1-V_0}{S} \vec{e_z} \Rightarrow  a = \frac{q(V_1-V_0)}{Sm}.
\end{align}
\end{linenomath*}
The ideal switch time computes then to 
\begin{linenomath*}
\begin{align}
    t_s = \frac{l_0\sqrt{m}}{\sqrt{2T_0}} +  \frac{S\sqrt{2m}}{ \sqrt{T_1} + \sqrt{T_0}} + \frac{l_1\sqrt{m}}{\sqrt{2T_1}}.
    \label{eq:eq:switch_time_dec_simpl}
\end{align}
\end{linenomath*}
Considering no energy modification ($T_0=T_1$) in the first gap, e.g., when the second gap is used for an energy modification, equation \ref{eq:eq:switch_time_dec_simpl} simplifies to
\begin{linenomath*}
\begin{align}
    t_{s,a} = \frac{\sqrt{m}}{\sqrt{2T_0}}(l_0 + S + l_1).
    \label{eq:switch_time_acc_simpl}
\end{align}
\end{linenomath*}
\subsubsection{Axially aligned cylindrical electrodes}
A more realistic potential in a gap between two axially aligned cylindrical electrodes was computed, e.g., in reference \cite{F_H_Read_1971}, from which follows the electrical field and consequently the acceleration, which both depend on the position $z$ of the ion bunch, where $z = 0$ is in the center of the gap: 
\begin{linenomath*}
\begin{align}
    \vec{E(z)} = - \nabla \phi(z) = \frac{V_1-V_0}{2S}(\tanh{(w(z+S/2)} + \tanh{(w(z-S/2))}) \vec{e_z} \Rightarrow \vec{a} = \frac{q|\vec{E}|}{m}, 
\end{align}
\end{linenomath*}
where $w$\,=\,2.64/$D$. By integration, the ideal switch time $t_s$ can be determined
\begin{linenomath*}
\begin{align}
    t_s = & \int^{l_1+S/2}_{-(l_0+S/2)}\left(\frac{q(V_1-V_0)}{wSm} \ln\left( \frac{\cosh(w(z+S/2))}{\cosh(w(z-S/2))} \right) + \frac{2T_0}{m}\right)^{-\frac{1}{2}} dz,
    \label{eq:switch_time}
\end{align}
\end{linenomath*}
which can be solved numerically, and accounts for both acceleration and deceleration using first or second gap. Using the second gap, which implies $V_0$\,=\,$V_1$, eq. (\ref{eq:switch_time}) simplifies to eq. (\ref{eq:switch_time_acc_simpl}).
\subsection{Length of drift tube and radius}
\begin{figure}
    \centering
    \includegraphics[width=1\textwidth]{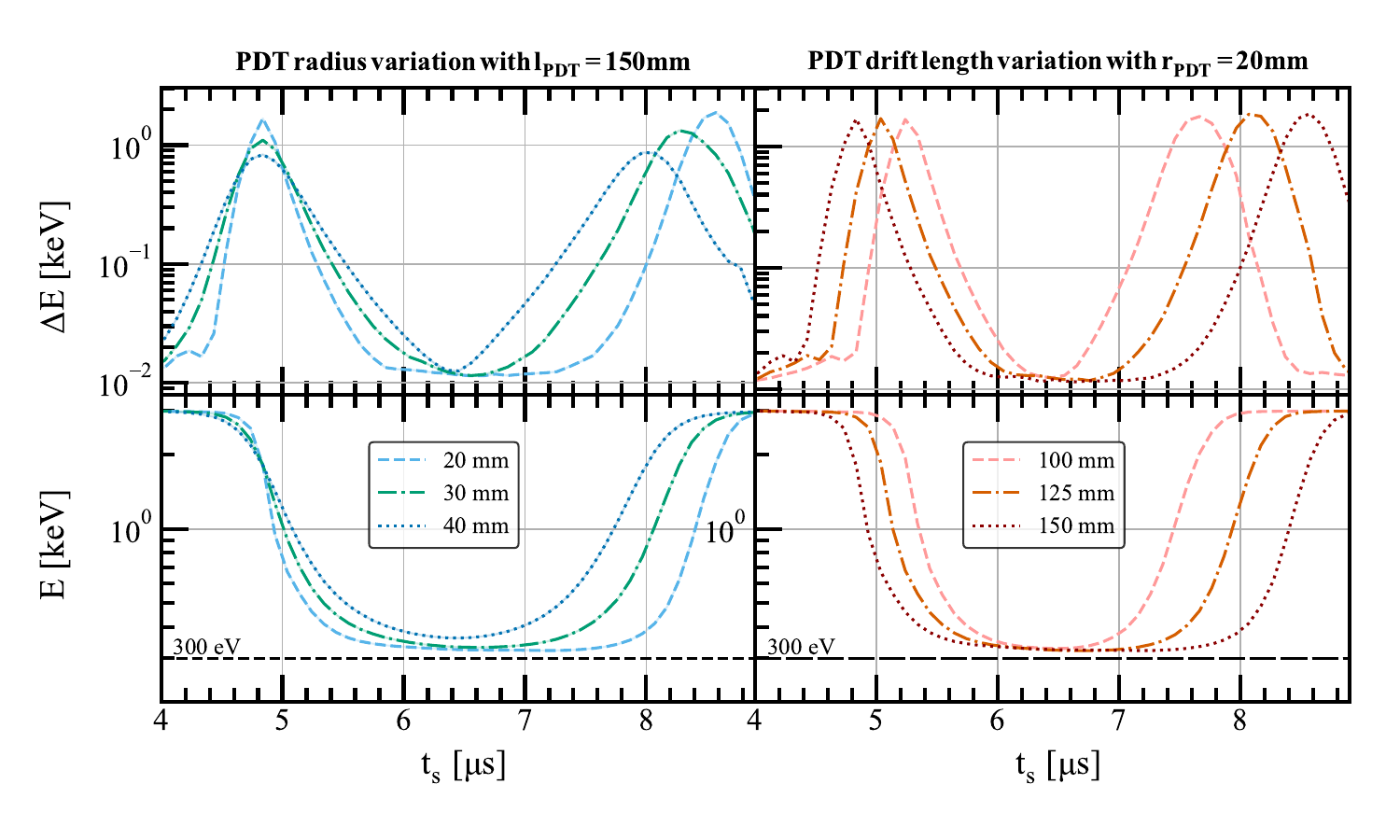}
    \caption{The resulting energy $E$ and the energy spread $\Delta E$ of the deceleration as a plot of the switch time $t_s$ depending on PDT radius $r_d$ (left) and length $l_d$ (right). A $^{40}$Ar$^+$ ion bunch with a bunch length of \SI{300}{\nano\second} and an emittance of 8.3\,$\pi$\,mm\,mrad was simulated to determine the preferable dimensions. }
    \label{fig:geometry_pdt}
\end{figure}

The PDT was characterized by simulating the deceleration of $^{40}$Ar$^+$ ions with a kinetic energy of \SI{3}{\kilo\electronvolt} ions to \SI{0.3}{\kilo\electronvolt}, which is required for ion trapping in the RFQcb of the PUMA ion source beamline. The ions have, as above, a bunch length of \SI{300}{\nano\second} and an emittance of 8.3\,$\pi$\,mm\,mrad. The length of the drift tube $l_d$ and the radius of the PDT $r_d$ have an impact on the energy modification efficiency and longitudinal acceptance as shown in \ref{fig:geometry_pdt}. The larger $r_d$, the smaller is the switch-time plateau, on which the desired energy $E$ with energy spread $\Delta E$ is reached. This is caused by the fringe field leaking further into the drift region for larger $r_d$. Thus, a radius of \SI{20}{\milli\meter} was chosen considering longitudinal acceptance and plateau length. Consequently, $l_d$\,=\,\SI{150}{\milli\meter} was chosen which is sufficient for accepting bunches with a bunch length < \SI{1.4}{\micro\second} in the case the deceleration from \SI{3}{\kilo\electronvolt} ions to \SI{0.3}{\kilo\electronvolt}.
\subsection{Deceleration of ions}
\begin{figure}
    \centering
    \includegraphics[width=0.8\textwidth]{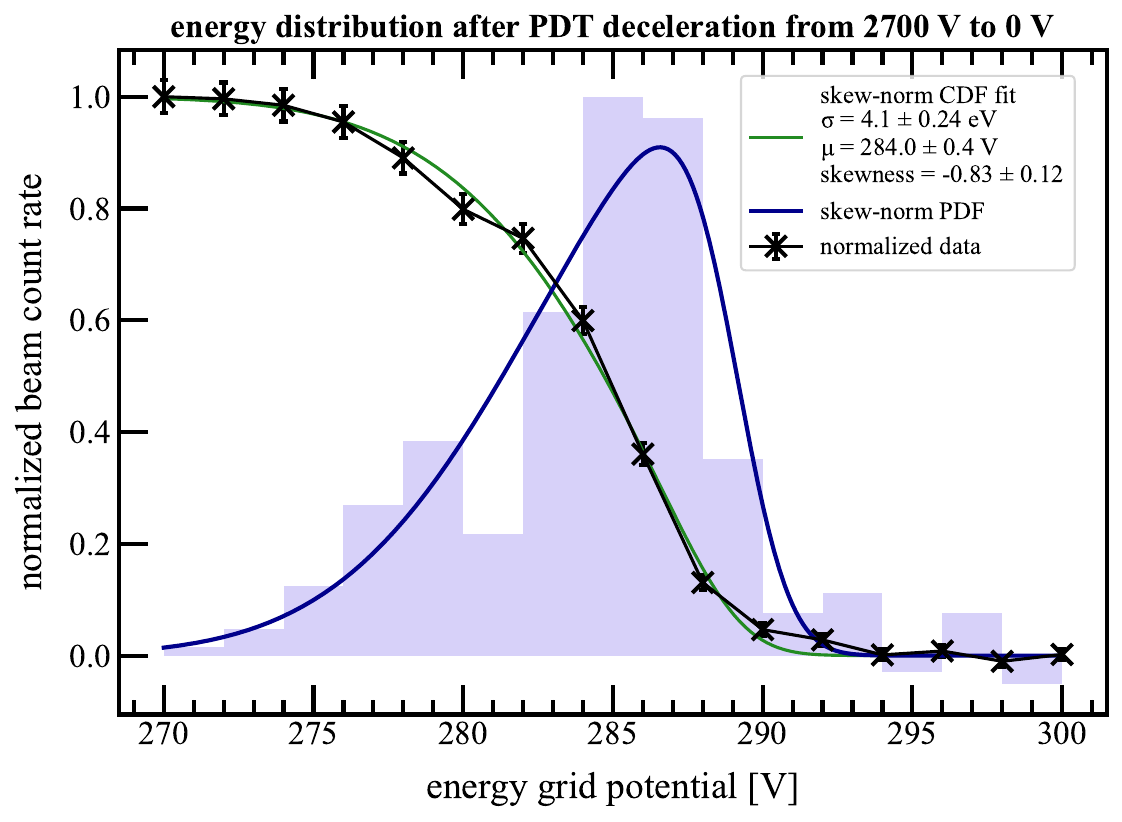}
    \caption{The beam energy distribution was measured using an energy analyser grid. By fitting the cumulative distribution function (CDF) the mean energy and its standard deviation could be determined using a skew normal distribution fit. In the background the difference between two measurement points is plotted together with the skew-now PDT to further visualize the shape of the distribution.}
    \label{fig:energy}
\end{figure}

\begin{figure}
    \centering
    \includegraphics[width=1\textwidth]{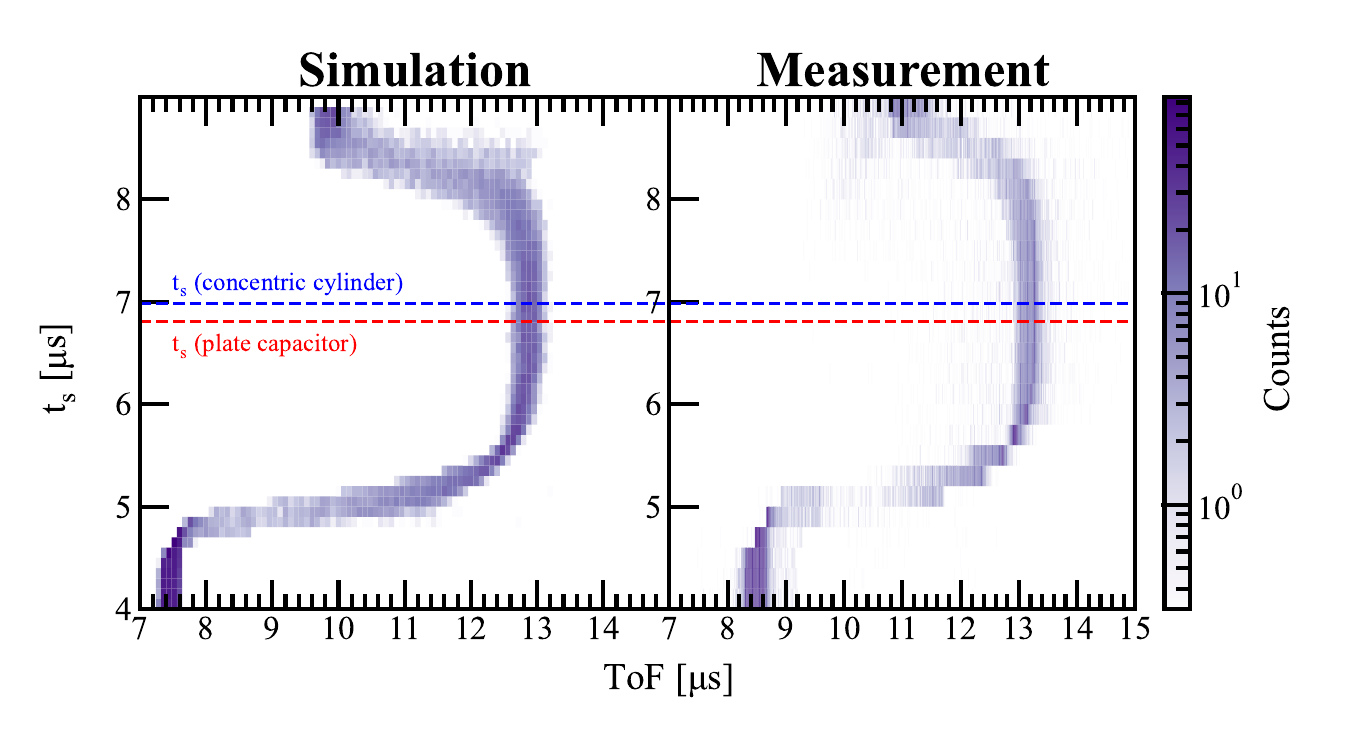}
    \caption{Plot of the ion ToF spectrum as a function of the switch time $t_s$ of the PDT for a deceleration from \SI{3}{\kilo\electronvolt} to \SI{0.3}{\kilo\electronvolt}, comparing the simulation (left) to the measurement (right). The intensity is color coded and normalized to the total count rate per slice.}
    \label{fig:deceleration}
\end{figure}

\begin{figure}
    \centering
    \includegraphics[width=0.8\textwidth]{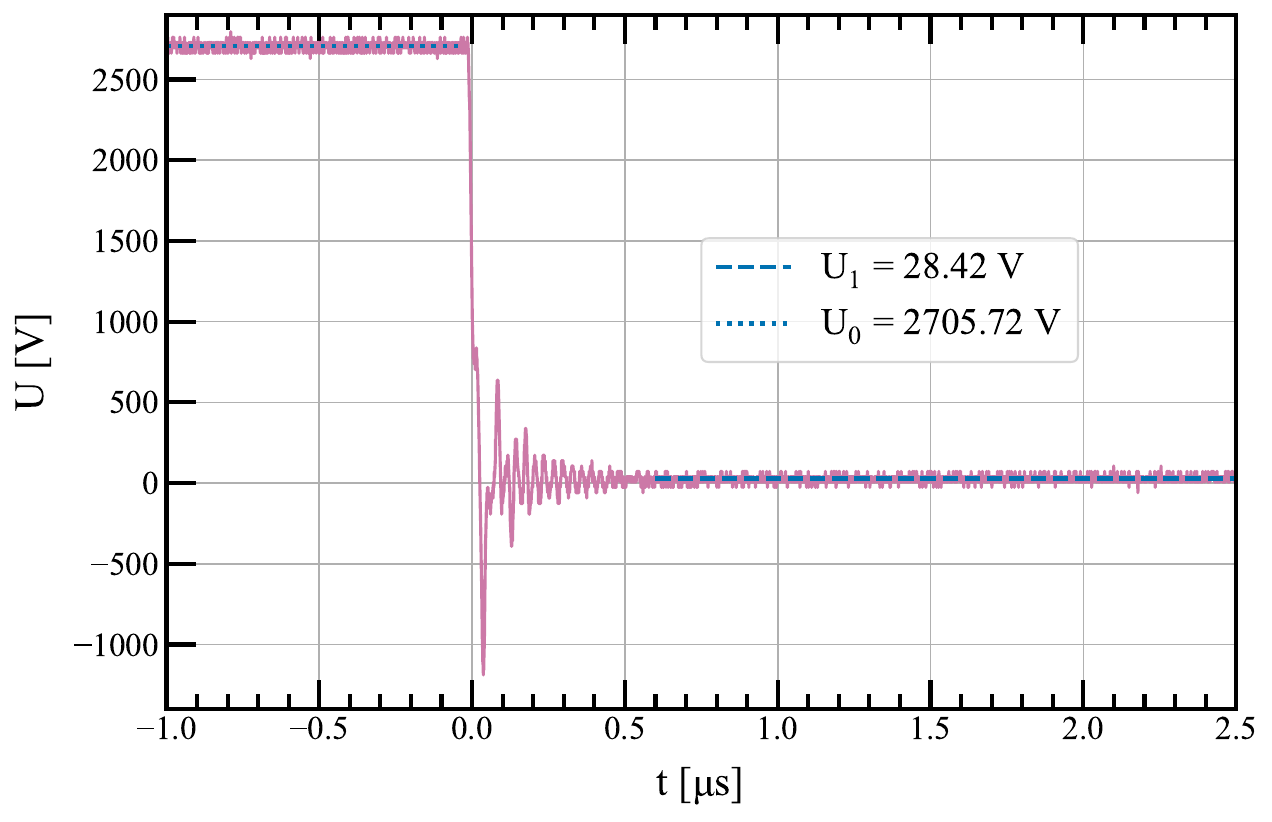}
    \caption{A measurement of the output voltage of the PDT switch over the switch time. When switching from \SI{2.7}{\kilo\volt} to ground, a damped oscillation is visible over a time of about \SI{0.5}{\micro\second} until the applied potential converges at \SI{28.42}{\volt}. This increases the energy that is reached after deceleration. }
    \label{fig:switch_pattern}
\end{figure}

In figure \ref{fig:deceleration} a ToF spectrum for an increasing switch time $t_s$ is displayed for the simulation (left) and the measurement (right). The results are normalized to total counts per scan step. The measurement was performed with PDT 1 in section I of the ion source beamline, figure \ref{fig:beamline_ion_source}. Ions with the same properties as in the simulation are produced in the ion source and their ToF is measured after a deceleration from \SI{3}{\kilo\electronvolt} to \SI{0.3}{\kilo\electronvolt}. The horizontal lines in red  (eq. (\ref{eq:eq:switch_time_dec_simpl})) and blue (eq. (\ref{eq:switch_time})) show the results of the calculation of the ideal switch time for the dimensions of the beamline. 
\paragraph{Simulation}
In the simulation an energy of \SI{322.4}{}\,$\pm$\,\SI{5.2}{\electronvolt} was reached with a switch plateau of approximately \SI{1.4}{}\,$\pm$\,\SI{0.1}{\micro\second}. Applying \SI{2.7}{\kilo\volt} to the PDT, one would expect to reach a final kinetic energy of \SI{300}{\electronvolt}, but the deviation can be explained by the behaviour of the PDT switch. The switch pattern of the PDT switch (Behlke HTS 81-06-GSM) was recorded and implemented in the simulations (see figure \ref{fig:switch_pattern}). The initial measured voltage applied to the PDT was \SI{2705.72}{\volt}. Switching induces significant noise during and shortly after the switch down, afterwards a damped oscillation converges to \SI{28.42}{\volt}, leading to a final energy of \SI{322.7}{\electronvolt}.
\paragraph{Measurement}
In the measurement, the $^{40}$Ar$^+$ ToF spectra show a similar behavior as in the simulation. The switch plateau length is determined to be \SI{1.6}{}\,$\pm$\,\SI{0.1}{\micro\second}. The measured ToF spectrum is shifted on the ToF axis, because the ion production and thus the trigger for the data recording takes place at an unknown position further inside the ion source while in the simulations the ejection point of the ion source was considered as the starting position. An energy analyzer was used in between the PDT and the ToF detector to estimate the resulting energy. Three grids with a transmission of 92 -- 95 \% are axially aligned, and a potential is applied to the central grid to form a blocking potential. The measured ion count rate as a function of the blocking potential can be fitted with the cumulative distribution function (CDF) of a skew normal distribution, which allows to determine the mean energy $\mu$ and the standard deviation $\sigma$. For further visualization, the difference in ion counts between two points is plotted together with the corresponding skew normal PDF using the parameters from the CDF, as seen in figure \ref{fig:energy}. The skew normal distribution has the form $f(x) = 2 \phi(x) \Phi(\alpha x)$, with $\phi(x)$ as the standard normal probability density distribution function and $\Phi(\alpha x)$ as the standard normal cumulative distribution function with skewness parameter $\alpha$. \\
In this way, the mean energy was determined to be \SI{284.0}{}\,$\pm$\,\SI{0.4}{\electronvolt}. This is about \SI{38}{\electronvolt} below the expected value, explaining the deviations in ToF difference of simulation and measurement. Recording the switch pattern again after the measurement, it was observed that by small modifications of the switch setup, e.g., the electrical connection, the maximum applied current or the type of grounding, the voltage when switching to \SI{0}{V} varies by several tens of volts. This is effect is less pronounced when applying a non-zero voltage to the switch channel, and could probably be prevented by grounding the channel instead of connecting it to a power supply. \\
Furthermore, several deceleration schemes were tested with the setup to modify the ion energy from \SI{3}{\kilo\electronvolt} to \SI{0.3}{\kilo\electronvolt} or \SI{1}{\kilo\electronvolt}. The results are summarized in table \ref{tab:dec_results}. Acceleration schemes were successfully tested for a re-acceleration after the RFQcb to \SI{4}{\kilo\electronvolt} but are not presented in the table due to the voltage limitation on the energy analyzer.
\begin{table}
\caption{Measured ion bunch parameters after deceleration with the PDT for different switch patterns.}
\centering
\begin{tabular}{c|c|c|c|c}
\centering
Switch pattern & E [eV]       & $\sigma_E$ [eV] & Skewness $\alpha$ & TOF FWHM [ns] \\ \hline \hline
2700\,V to 0\,V  & 284.0 $\pm$ 0.4  & 4.1 $\pm$ 0.2          & $-$0.83 $\pm$ 0.12     & 294 $\pm$ 51  \\
0\,V to $-$2700\,V    & 322.9 $\pm$ 0.8     & 7.4 $\pm$ 0.5          & $-$0.9 $\pm$ 0.1    & 277 $\pm$ 12   \\
2500\,V to $-$200\,V  & 327.6 $\pm$ 2.1  & 9.3 $\pm$ 1.3          & 0.3 $\pm$ 0.2     & 302 $\pm$ 34         \\
200\,V to $-$2500\,V  & 317.8 $\pm$ 0.6  & 6.3 $\pm$ 0.4          & $-$0.7 $\pm$ 0.1    & 264 $\pm$ 16 \\
2000\,V to 0\,V     & 988.4 $\pm$ 0.9       & 8.8 $\pm$ 0.5          & $-$1 $\pm$ 0.1      & 279 $\pm$ 32         \\ 
0\,V to $-$2000\,V    & 1019.5 $\pm$ 0.6 & 6.0 $\pm$ 0.4          & $-$0.9 $\pm$ 0.1    & 279 $\pm$ 18         \\
1000\,V to $-$1000\,V  & 1015.1 $\pm$ 0.7    & 5.9 $\pm$ 0.5          & $-$0.7 $\pm$ 0.2    & 278 $\pm$ 21        
\end{tabular}
\label{tab:dec_results}
\end{table}
\subsection{Longitudinal acceptance}
\begin{figure}
    \centering
    \includegraphics[width=1\textwidth]{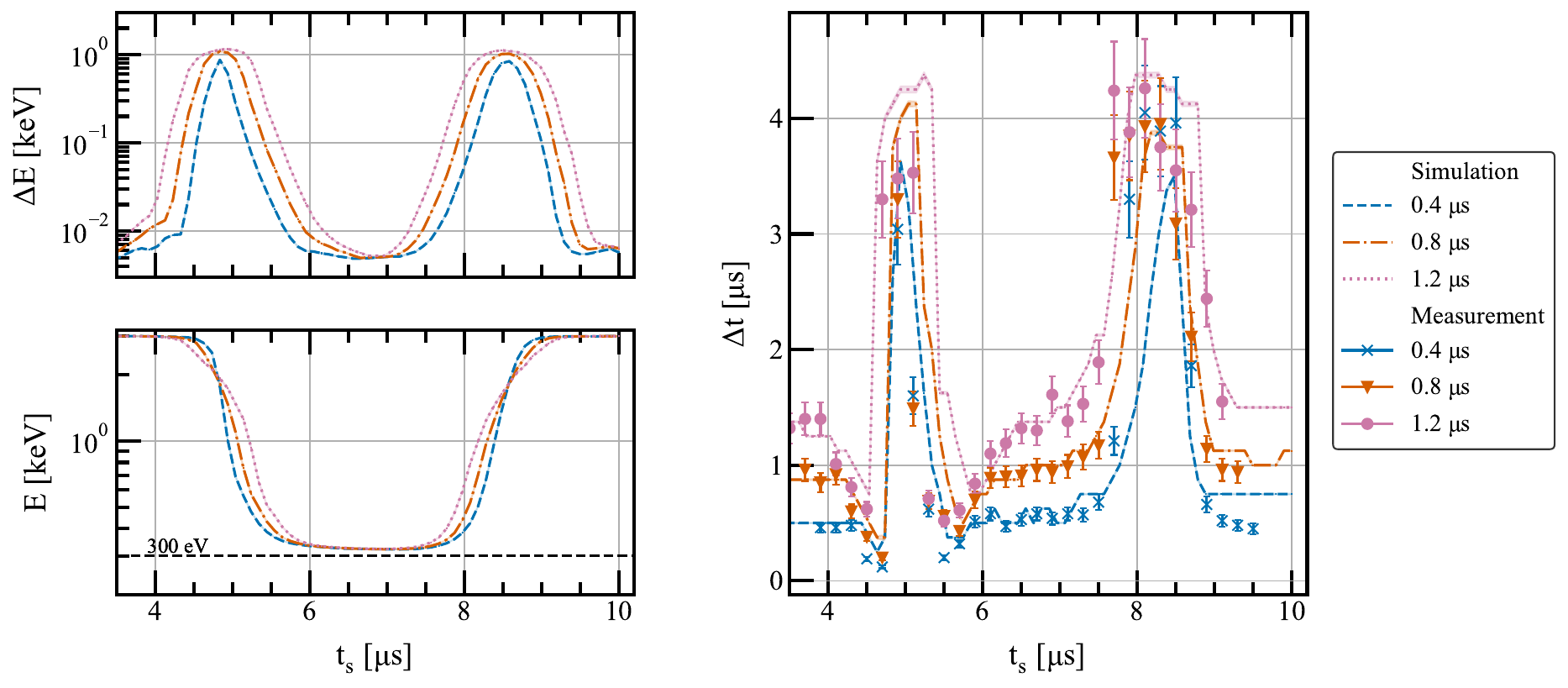}
    \caption{The energy $E$ and energy spread $\Delta E$ (left) and bunch length $\Delta t$ (right) depending on the switch time $t_s$ of the PDT for different bunch lengths 0.4, 0.8 and 1.2\,$\mu$s for the deceleration of $^{40}$Ar$^+$ ions from \SI{3}{\kilo\electronvolt} to \SI{0.3}{\kilo\electronvolt}.}
    \label{fig:bunch_width}
\end{figure}

In figure \ref{fig:bunch_width} (left), the mean ion energy is plotted as a function of $t_s$ for uniform intial bunch lengths of \SI{0.4}{}, \SI{0.8}{} and \SI{1.2}{\micro\second}. For increasing bunch lengths, larger parts of the bunch are in the fringe field or partly outside the drift tube when switching at non-ideal switch times, and thus the switch plateau for minimal final energy and energy spread decreases. The initial energy spread is \SI{10}{\electronvolt}, which is improved for some $t_s$ for all tested bunch widths. \\
On the right side of figure \ref{fig:bunch_width}, the resulting bunch lengths are plotted with respect to $t_s$. Measurement (markers) and simulation (dashed lines) can be compared. For the measurements, the bunch length could be changed using a pulsed deflector electrode in front of the ion source to chop the continuous beam, resulting in a uniform bunch shape. Since the shape of the ion bunch distribution after switching can be non-uniform and non-Gaussian, depending on $t_s$ the bunch length was determined independently of the ion distribution within the ion bunch for a better comparison. For this, the width is defined by the difference of the ToF positions at which the ion distribution falls below and rises above 10\,\% of the distributions maximum value. The simulations describe the measurement results well, only the bunch length of the bunches located in the fringe field while switching is slightly overestimated. It can be concluded that by choosing an appropriate $t_s$, the bunch length degradation can be minimized. This is especially important in cases where long distances have to be covered after an energy modification. 
\section{Iris shutter assembly}
\subsection{Description}
\label{sec:iris_descr}
The dimensions of the iris shutter assembly, which is made of a C-276 alloy (SAHM, 95032), can be found in appendix \ref{drw:iris}. Figure  \ref{fig:iris_photo} shows a photograph of the assembly. The iris shutter is mounted on a customized double-sided CF160 flange with two radial ports, one of which holds a micrometer precision linear feedthrough (VAb, NC16-100) that allows to regulate the opening of the iris aperture. The micrometer feedthrough is connected to the iris via a mechanical lever. With this, the iris aperture can be opened (\SI{30}{\milli\meter} diameter) or closed with 48 rough adjustment steps, which follow an almost linear dependency. Therefore, the iris diameter can be varied in steps of \SI{625}{\micro\meter} or better. In the case of the PUMA ion source beamline the shutters are meant to decouple the vacuum regions with different pressure of a beamline by reducing the conductance between the sections while still allowing a focused beam to pass through. \\ 
The iris apertures can additionally be used to optimize the overlap of ion beam velocity axis with the central axis, used for beamline tuning. The same technique could also be considered for laser beam overlap. By gradually closing it, the beams centrality and its size of focus can be checked, as described, e.g., in reference \cite{Iris_Zhang}. Furthermore, if the iris shutter is insulated from the flange by a thin teflon layer, one can electrically connect it to the second radial port and read the current of the impinging beam. 
\begin{figure}
    \centering
    \includegraphics[width=0.5\textwidth]{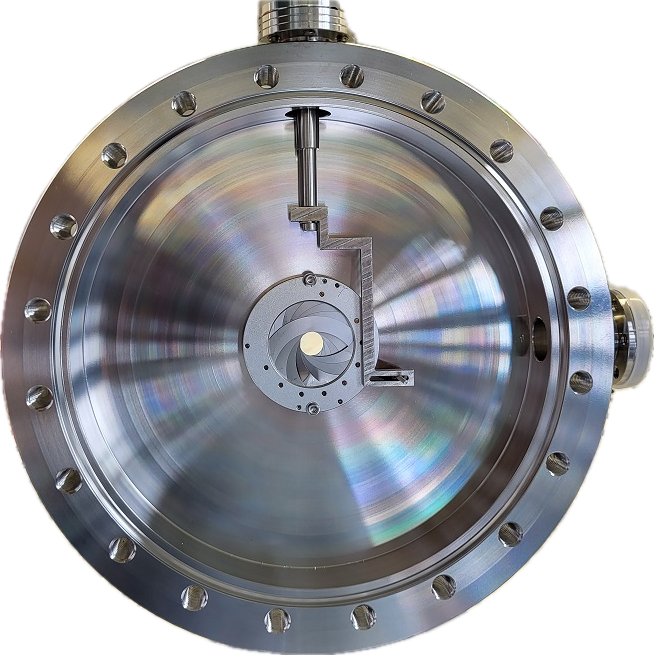}
    \caption{A photograph of the iris shutter assembly. The iris shutter can be positioned between two beamline sections to lower the pass-through area for residual gas particles. It can be opened and closed from outside the vacuum using a the linear feedthrough.}
    \label{fig:iris_photo}
\end{figure}
\subsection{Pressure between vacuum sections}
\begin{figure}[h]
    \centering
    \includegraphics[width=1\textwidth]{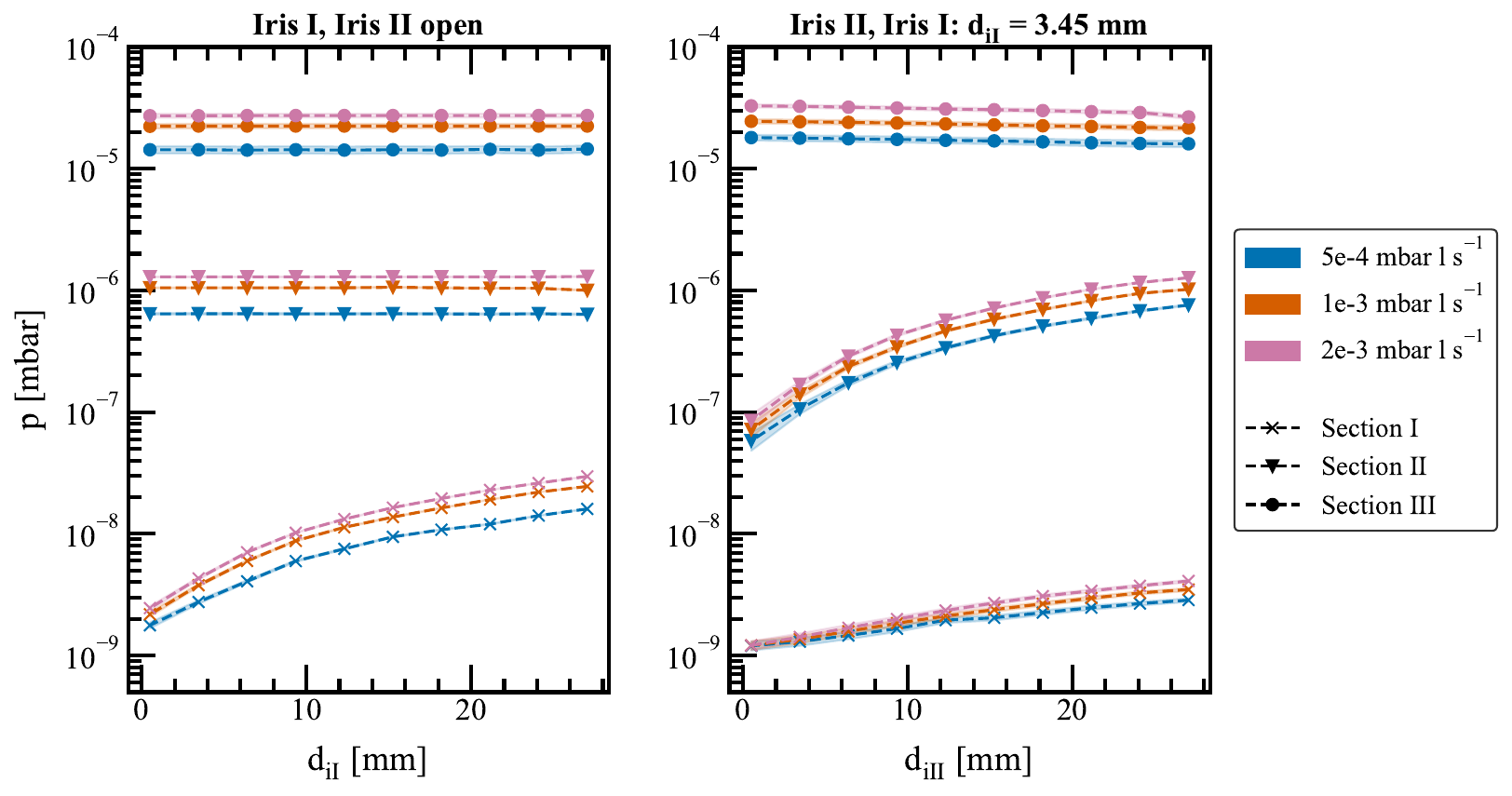}
    \caption{The measured pressure in section I, II and III depending on the opening diameter $d_i$ of the iris-type apertures. On the left site, the Iris I is varied while Iris II is kept open. The impact of Iris II is shown on the right side, while Iris I is kept on \SI{3.45}{\milli\meter}. Three buffer gas flow rates were considered: $5\cdot$10$^{-4}$\,mbar\,l\,s$^{-1}$ (red), $1\cdot$10$^{-3}$\,mbar\,l\,s$^{-1}$ (blue), $2\cdot$10$^{-3}$\,mbar\,l\,s$^{-1}$ (green).  }
    \label{fig:iris_shutter_measurement}
\end{figure}

In the PUMA ion source beamline, four iris apertures are installed, see Figure 1. Argon buffer gas is injected in the RFQcb section (section III), leading to a vacuum from 1$\cdot$10$^{-5}$\,mbar to 5$\cdot$10$^{-5}$\,mbar. Especially the antiproton beam line has to be shielded from a vacuum contamination to keep the vacuum < 10$^{-9}$\,mbar at the quadrupole bender junction. In the following, the impact of the apertures on the vacuum in sections I and II (see figure \ref{fig:beamline_ion_source}) is demonstrated. \\
In figure \ref{fig:iris_shutter_measurement}, three argon buffer gas flow rates were considered: $5\cdot$10$^{-4}$\,mbar\,l\,s$^{-1}$, $1\cdot$10$^{-3}$\,mbar\,l\,s$^{-1}$, $2\cdot$10$^{-3}$\,mbar\,l\,s$^{-1}$. Each section is differentially pumped with turbomolecular pumps. The pressure was measured for different iris aperture diameters $d_{\text{I}}$, $d_{\text{II}}$ and $d_{\text{III}}$ with full range and cold cathode gauges at the side flanges.In the left figure, the diameter of Iris I is varied while Iris II is kept open. Reducing the opening of Iris I results in a decrease of the pressure by one order of magnitude. When also closing Iris II, a similar drop in pressure can be observed in section II (right). Meanwhile, the pressure in section I decreased by another factor two, when keeping Iris I at a diameter of \SI{3.45}{\milli\meter}. Gas-flow simulations with the software MolFlow+ \cite{molfow} were carried out to test the effect of using a stack of multiple iris shutters in a row. However, the non-evacuated region between the shutters acts as buffer zone for higher pressure and overall deteriorates the vacuum level instead of improving it. If the vacuum has to be further improved, an implementation of ion getter pumps and the use of baking is required.
\subsection{Outgassing rates}
The iris shutters were baked up to temperatures of 250\,°C in a test campaign to determine their outgassing rates. 48 hours after the end of a 72 hour bake-out campaign, the iris shutter had a hydrogen degassing rate of less than 1.1$\cdot10^{-9}$\,mbar\,l\,s$^{-1}$s. Furthermore, elements with atomic mass units of 53, 64, 66 and 68 are present after the bakeout, but with a factor 1000 less than hydrogen. \cite{CERN_vacuum_report}.
\section{Conclusion}
For the PUMA ion source beamline, operating with voltages up to \SI{5}{\kilo\volt}, three ion-optical components have been designed for an adequate transport of the beam towards the PUMA antiproton beamline, as well as for improving vacuum levels between the buffer gas insertion point for the RFQcb and vacuum-sensitive parts of the beamline. The comparision to simple models and simulations have been shown to be predictive and could be reliably used for a setup design.\\
The einzel lens and steerer assembly successfully focuses and steers the beam for the injection into the RFQcb, which has the smallest acceptance in the beamline. Furthermore, the PDT was used to decelerate ions to energies as low as \SI{0.3}{\kilo\electronvolt}, while allowing to minimize the energy spread and bunch length by choosing appropriate switch times. Re-acceleration from \SI{0.3}{\kilo\electronvolt} to \SI{4}{\kilo\electronvolt} after the RFQcb was tested accordingly. The iris shutters significantly limit the impact of the RFQcb buffer-gas injection on the other vacuum sections by one order of magnitude per section shielded with the iris shutter assembly
\acknowledgments
PUMA is funded by the European Research Council through the ERC grant 
PUMA-726276 and the Alexander-von-Humboldt foundation. The development of PUMA and its implementation at CERN are supported by the TU Darmstadt, Germany and CERN, Switzerland. Furthermore, this work has been sponsored by the Wolfgang Gentner Programme of the German Federal Ministry of Education and Research (grant no.13E18CHA). Additionally, we want to thank the CERN vacuum team (TE-VSC-IVO) for performing the measurement of the iris shutter outgassing rates. 
\subsection*{Data Availability Statement}
Data are available on request from the authors. 
%\begin{thebibliography}{99}

%\bibitem{hinterberger}
%F. Hinterberger,
%\emph{Ion optics with electrostatic lenses
%},\emph{J. Abbrev.} {\bf vol} (2006) pg. 27-44

%\bibitem{b}
%Author,
%\emph{T},
%arxiv:1234.5678.

%\bibitem{c}
%Author,
%\emph{Title},
%Publisher (year).

%\end{thebibliography}
%\printbibliography

%\begin{appendices}
\appendix
\section{Drawings}

\includepdf[scale=0.75,pages=1,pagecommand=\subsection{Einzel lens and steerer assembly}\label{drw:lens}, noautoscale,offset= 72.5 -75]{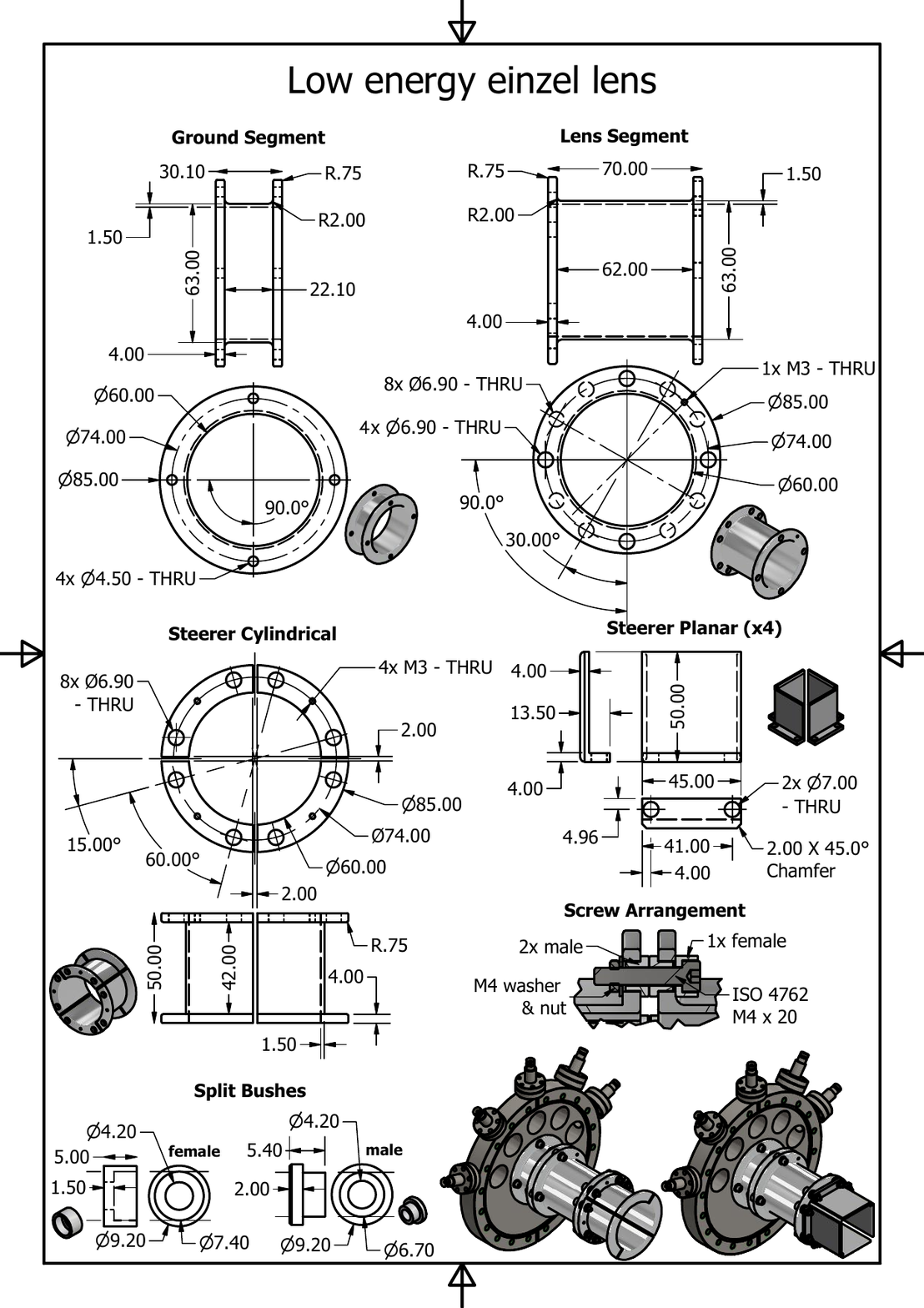}
\includepdf[scale=0.75,pages=1,pagecommand=\subsection{Mounting flange}\label{drw:flange}, noautoscale,offset= 72.5 -75]{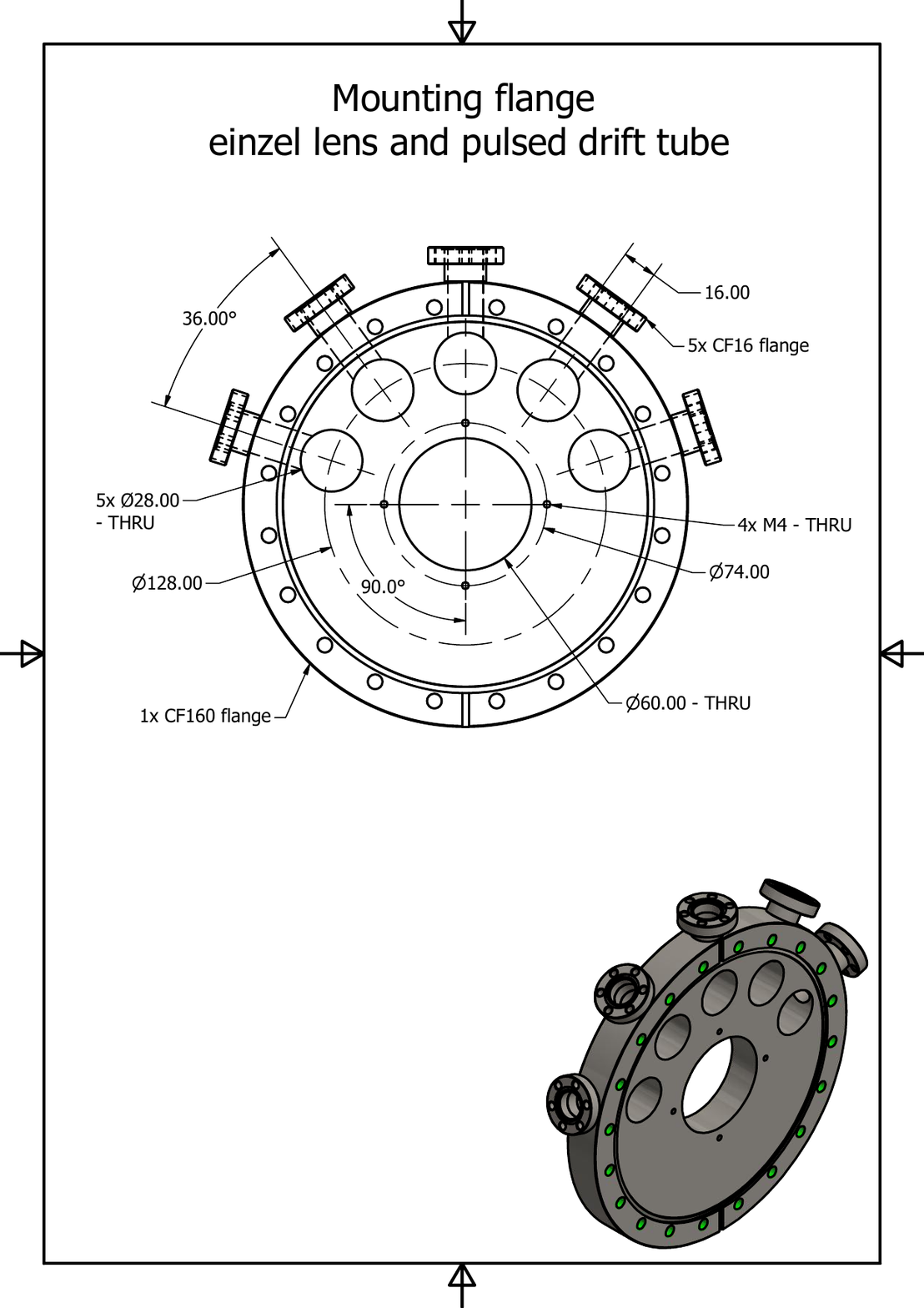}
\includepdf[scale=0.75,pages=1,pagecommand=\subsection{Pulsed drift tube}\label{drw:pdt}, noautoscale,offset= 72.5 -75]{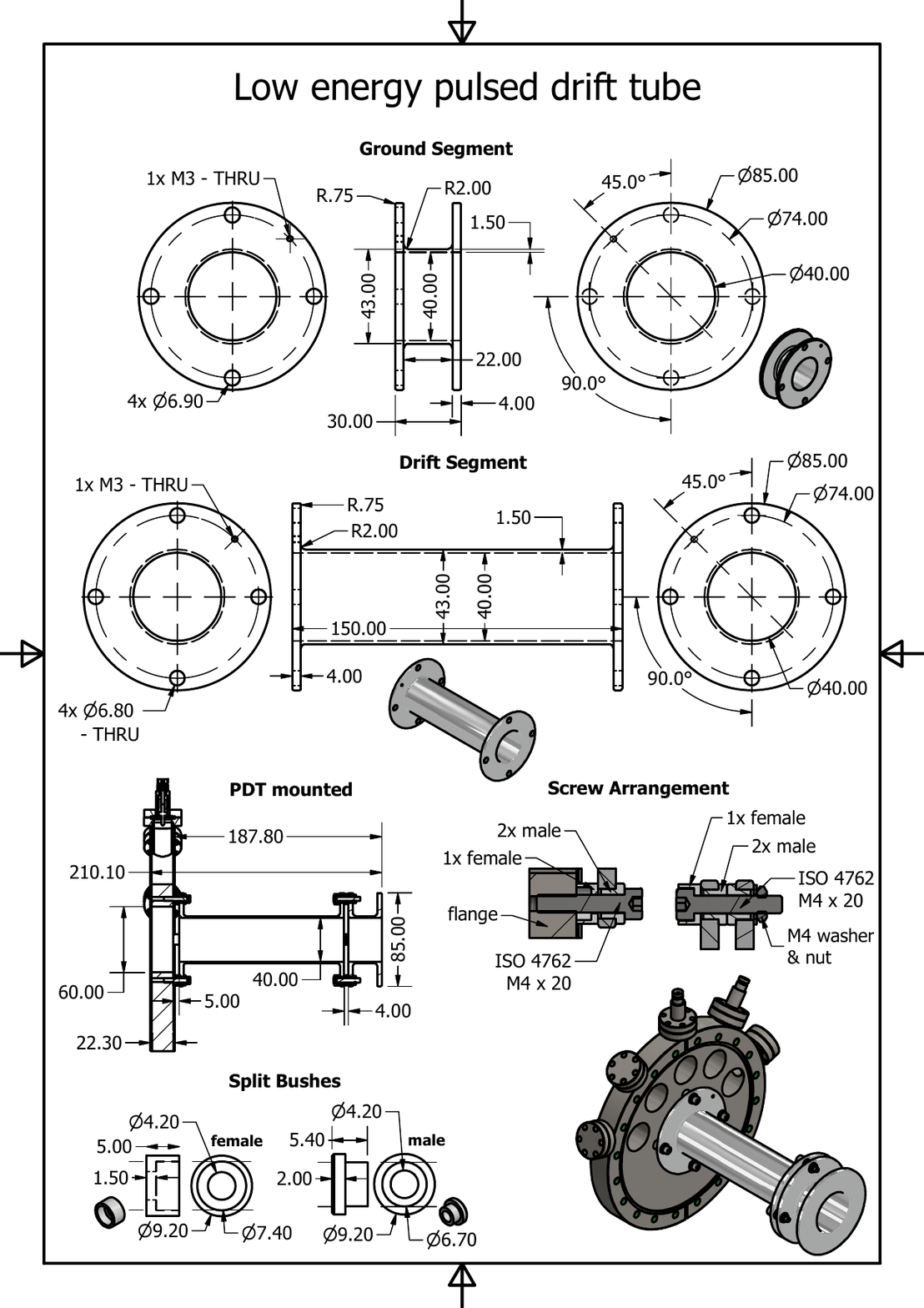}
\includepdf[scale=0.75,pages=1,pagecommand=\subsection{Iris shutter assembly}\label{drw:iris}, noautoscale,offset= 72.5 -75]{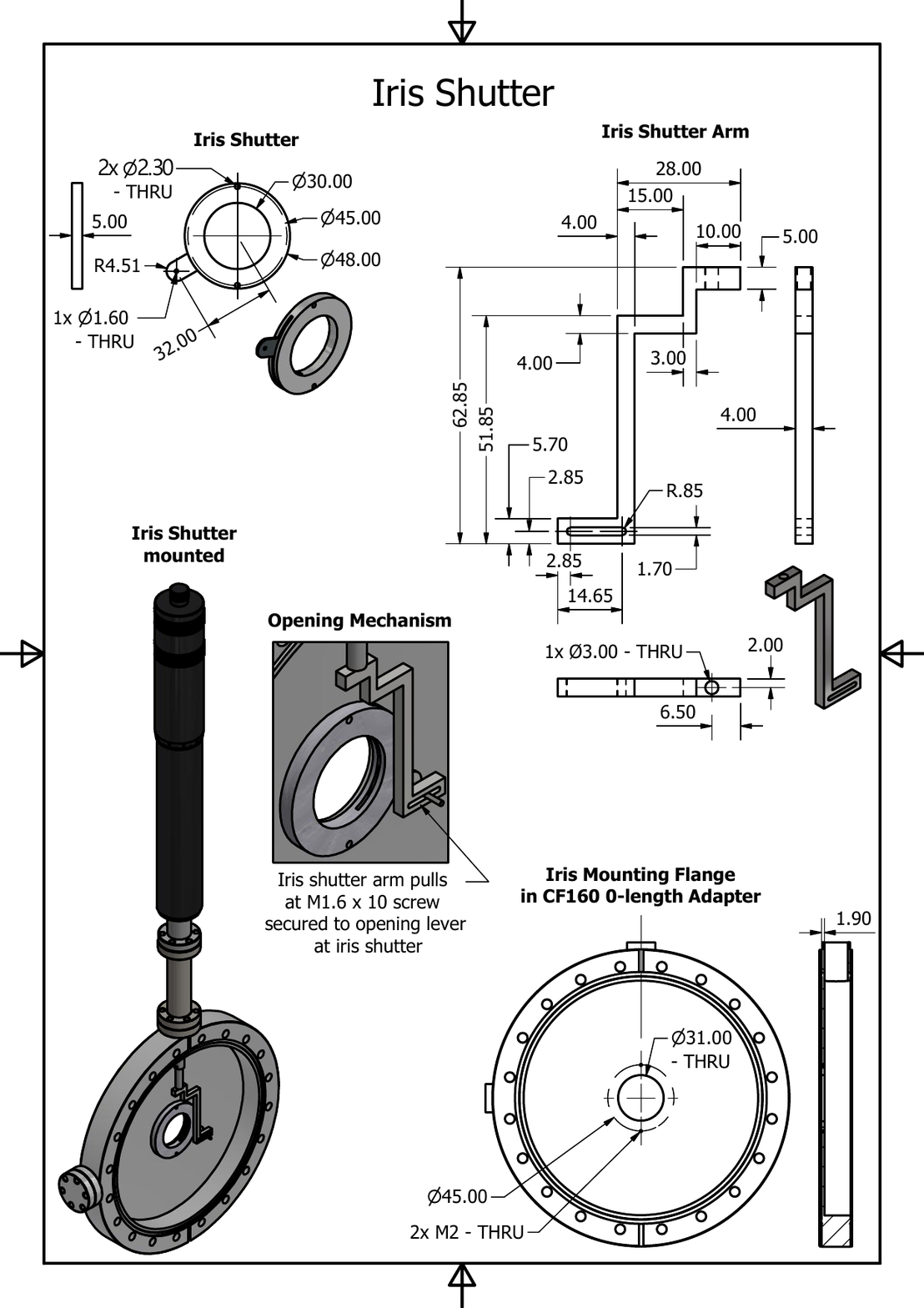}

%\end{appendices}

\bibliographystyle{JHEP} % or try abbrvnat or unsrtnat
\bibliography{biblio} % refers to example.bib

\end{document}